\documentclass[showpacs,aps,prl,twocolumn]{revtex4}
\usepackage{dcolumn}
\usepackage{graphicx}
\usepackage{amsmath}
\usepackage{latexsym}
\usepackage{amssymb}

\begin{document}
\title{Quantification of quantum entanglement in a multiparticle system of two-level atoms interacting with a squeezed vacuum state of the radiation field}
\author{Ram Narayan Deb}
\address{Department of Physics, Chandernagore College,
 Chandernagore, Hooghly, Pin-712136, West Bengal, India}

\begin{abstract}
We quantify multiparticle quantum entanglement in a system of $N$ two-level atoms interacting with a squeezed vacuum state of the electromagnetic field. We calculate the amount of quantum entanglement present among one hundred such two-level atoms and also show the variation of that entanglement with the radiation field parameter. We show the continuous variation of the amount of quantum entanglement as we continuously increase the number of atoms from $N = 2$ to $N = 100$. We also discuss that the multiparticle correlations among the $N$ two-level atoms are made up of all possible bipartite correlations among the  $N$ atoms. 
\end{abstract}

\maketitle

\section{I. Introduction}
The quest for the quantification of quantum entanglement in multiparticle systems is an important new arena of present day research. Over the past few years there has been a growing interest in studying quantum entanglement in multiparticle systems
\cite{Hald}-\cite{Ram}. Peres and Horodecki \cite{Peres}, \cite{Horodecki} made an important step toward the understanding of quantum entanglement in the context of bipartite states. In Ref. \cite{Paternostro}, the sufficient and necessary conditions to induce entanglement on two remote qubits, by means of their respective linear interaction, with a two-mode driving field have been studied. A lot of work has also been done to understand the relationship between quantum entanglement and spin squeezing \cite{Sorensen}-\cite{Dalton}. 
In Ref. \cite{Giuseppe}, a complete set of generalised spin squeezing inequalities for detecting entanglement in an ensemble of qudits have been presented. It has also been shown, how to detect $k$-particle entanglement and bound entanglement. 
In this paper, we quantify quantum entanglement present in a system of $N$ two-level atoms in interaction with the squeezed vacuum state of the radiation field. We also discuss that the multiparticle correlations among the $N$ two-level atoms are made up of all possible bipartite correlations among the  
$N$ atoms. We also notice that the system shows high value of quantum entanglement, when there is no spin squeezing, as defined in Ref. \cite{Kitagawa}, in the system.

 In Ref. \cite{Ram}, we proposed the necessary and sufficient condition for the presence of quantum entanglement in arbitrary symmetric pure states of $N$ two-level atoms. We introduced a parameter, called quantum entanglement parameter, to detect and quantify quantum entanglement in such multiatomic systems. In this paper, we use that parameter to quantify quantum entanglement in a real physical system. The physical system, that we take, is a system of $N$ two-level atoms interacting with the squeezed vacuum state of the electromagnetic field. This system is of considerable interest in Quantum Optics.
We quantify the amount of quantum entanglement among the atoms of this system for the number of atoms, varying from $N=2$ to $N=100$. We also show that the system of atoms shows high quantum entanglement when there is no spin squeezing, as defined in Ref. \cite{Kitagawa},  in the system.
We emphasize the fact that, the quantum entanglement in such multiatomic systems can be quantified experimentally.

In section II, we present the construction of the quantum entanglement parameter. In section III, we present our study on quantum entanglement of $N$ two-level atoms interacting with the squeezed vacuum state of the radiation field. In section IV, 
we present the summary and conclusion.

\section{II. Construction of quantum entanglement parameter}

An atom has many electronic energy levels, but when it interacts with an external monochromatic electromagnetic field, the atom makes a transition from one of its energy level to the other. In this case, we mainly concentrate on those two energy levels and hence the atom is called as a two-level atom.

We consider a system of $N$ such two-level atoms.
Now, if among the assembly of $N$ such two-level atoms, the $n$-th atom has the upper and lower energy levels, denoted as 
$|u_n\rangle$ and $|l_n\rangle$, respectively, then, we can construct the pseudo-spin operators (with $\hbar = 1$),
\begin{eqnarray}
\hat{J}_{n_x} &=& (1/2)\big(|u_n\rangle\langle l_n| + |l_n\rangle
\langle u_n|\big),\label{1.1a1}\\
\hat{J}_{n_y} &=& (-i/2)\big(|u_n\rangle\langle l_n| - |l_n\rangle
\langle u_n|\big),\label{1.1a2}\\
\hat{J}_{n_z} &=& 
(1/2)\big(|u_n\rangle\langle u_n| - |l_n\rangle\langle l_n|
\big),
\label{1.1a3}
\end{eqnarray}
such that,
\begin{equation}
 [\hat{J}_{n_x}, \hat{J}_{n_y}] = i\hat{J}_{n_z},
\label{1.2}
\end{equation}
 and two more relations with cyclic changes in $x$, $y$ and $z$. For the entire system of $N$ two-level atoms, we have collective pseudo-spin operators,
\begin{eqnarray}
\hat{J}_x = \sum_{i=1}^{N}\hat{J}_{i_{x}},~~~~ 
\hat{J}_y = \sum_{i=1}^{N}\hat{J}_{i_{y}},~~~~ 
\hat{J}_z = \sum_{i=1}^{N}\hat{J}_{i_{z}}.
\label{1.3}
\end{eqnarray}
The individual atomic operators satisfy
\begin{eqnarray}
\big[\hat{J}_{1_x}, \hat{J}_{2_y}\big] = 0,~
\big[\hat{J}_{1_x}, \hat{J}_{1_y}\big] = i\hat{J}_{1_z}, ~
\big[\hat{J}_{2_x}, \hat{J}_{2_y}\big] = i\hat{J}_{2_z},...
\label{1.2a1}
\end{eqnarray}
 As a consequence of these commutation
relations, the collective pseudo-spin operators 
$\hat{J}_x$, $\hat{J}_y$ and $\hat{J}_z$ satisfy,
\begin{equation}
[\hat{J}_x , \hat{J}_y] = i\hat{J}_z
\label{1.2a2}
\end{equation} 
and two more
relations with cyclic changes in $x$, $y$ and $z$.

The simultaneous eigenvectors of $\hat{J}^2 = \hat{J}_x^2
+ \hat{J}_y^2 + \hat{J}_z^2$ and $\hat{J}_z$ are denoted
as $|j,m\rangle$ where
\begin{equation} 
\hat{J}^2|j,m\rangle = j(j+1)|j,m\rangle
\label{1.2a3}
\end{equation}
 and 
\begin{equation}
\hat{J}_z|j,m\rangle = m|j,m\rangle.
\label{1.2a4}
\end{equation} 
The quantum number $j$ is related to the number of atoms 
$N$ as $j = N/2$ and $m = -j, -j+1, ....(j-1), j$.
The collective quantum state vector for a system of $N$
two-level atoms can be expressed as a linear superposition of $|j,m\rangle$ as  
\begin{equation}
|\psi_j\rangle = \sum_{m=-j}^{j}c_{m}|j,m\rangle.
\label{1.2a5}
\end{equation}
Now, to know whether an arbitrary quantum state vector $|\psi_j\rangle$ of the system is an atomic coherent state or atomic squeezed state, we calculate the mean pseudo-spin vector
\begin{equation}
\langle\hat{\mathbf{J}}\rangle = \langle\hat{J}_x\rangle
\hat{i} + \langle\hat{J}_y\rangle\hat{j} + \langle\hat{J}_z\rangle\hat{k},\label{1.4}
\end{equation}
where $\hat{i}$, $\hat{j}$ and $\hat{k}$ are the unit vectors along positive $x$, $y$ and $z$ axes respectively, and the average values in the above expression are to be calculated over the state $|\psi_j\rangle$.
The mean pseudo-spin vector 
$\langle\hat{\mathbf{J}}\rangle$ points in an arbitrary direction in space. Therefore, we conventionally rotate the coordinate system $\{x,y,z\}$ to
$\{x^\prime, y^\prime, z^\prime\}$, such that 
$\langle\hat{\bf{J}}\rangle$ points along the 
$z^\prime$ axis and calculate the quantum fluctuations in 
$\hat{J}_{x^\prime}$ and $\hat{J}_{y^\prime}$ for the state
$|\psi_j\rangle$. These quantum fluctuations are
\begin{equation}
\Delta {J_{x^\prime,y^\prime}}^2 = \langle\psi_j|{\hat{J}_{x^\prime,y^\prime}}^2|\psi_j\rangle - \langle\psi_j|\hat{J}_{x^\prime,y^\prime}|\psi_j\rangle^2.
\label{2.5}
\end{equation}
Now for an atomic coherent state, also called coherent spin state (CSS) in literature, \cite{Kitagawa}, the above quantum fluctuations are equal and they are,
\begin{equation}
\Delta {J_{x^\prime}}^2 = \Delta {J_{y^\prime}}^2 = \frac{j}{2} = \frac{N}{4},
\label{2.5a}
\end{equation}
and for an atomic squeezed state, also called squeezed spin state (SSS) \cite{Kitagawa},
\begin{equation}
\Delta {J_{x^\prime}}^2 ~~~ or,~~~ \Delta {J_{y^\prime}}^2 <  \frac{j}{2} = \frac{N}{4}.
\label{2.5b}
\end{equation}
It is to be mentioned here that
\begin{equation}
\Delta {J_{x^\prime}}\Delta {J_{y^\prime}} \ge \frac{N}{4},
\label{2.5b11}
\end{equation}
which is the Heisenberg's uncertainty principle.

A collective state vector $|\alpha\rangle$ for a system of two atoms is said to be quantum mechanically entangled  if 
$|\alpha\rangle$ cannot be expressed as a direct product
of the two individual atomic state vectors i.e.
\begin{equation}
|\alpha\rangle \ne |\alpha_1\rangle \otimes|\alpha_2\rangle,
\label{1.7}
\end{equation}
where $|\alpha_1\rangle$ and $|\alpha_2\rangle$ are the 
state vectors of the two individual atoms \cite{Nielsen}. 

We now proceed to construct the quantum entanglement parameter for a system of $N$ two-level atoms.

An arbitrary symmetric pure state for a system of $N$ two-level atoms in the
$\{ m_1, m_2, m_3,....m_N \}$ representation is given as

\begin{eqnarray}
|\Psi\rangle &=& G_1\bigg\vert\frac{1}{2},\frac{1}{2},....\frac{1}{2}
\bigg\rangle
+ \frac{G_2}{\sqrt{{}^NC_1}}\Bigg[\bigg\vert-\frac{1}{2},\frac{1}{2},\frac{1}{2},....\frac{1}{2}
\bigg\rangle \nonumber\\
&+& \bigg\vert\frac{1}{2},-\frac{1}{2},\frac{1}{2},....\frac{1}{2}
\bigg\rangle + \bigg\vert\frac{1}{2},\frac{1}{2},\frac{1}{2},....
-\frac{1}{2}\bigg\rangle\Bigg]\nonumber\\ 
&+& \frac{G_3}{\sqrt{{}^NC_2}}\Bigg[\bigg\vert-\frac{1}{2},-\frac{1}{2},\frac{1}{2},....\frac{1}{2}
\bigg\rangle \nonumber\\
&+& \bigg\vert-\frac{1}{2},\frac{1}{2},-\frac{1}{2},....\frac{1}{2}
\bigg\rangle + \bigg\vert\frac{1}{2},\frac{1}{2},....
-\frac{1}{2},-\frac{1}{2}\bigg\rangle\Bigg]\nonumber\\
&+& ............G_{N+1}\bigg\vert-\frac{1}{2},-\frac{1}{2},-
\frac{1}{2},....-\frac{1}{2}\bigg\rangle,
\label{2.5c}
\end{eqnarray}

where $G_1$, $G_2$, ..., $G_{N+1}$ are constants and
 ${}^N C_r$ is given as
\begin{equation}
{}^NC_r = \frac{N!}{r!(N-r)!}.
\label{2.5d}
\end{equation}

Now, the mean pseudo-spin vector $\langle\hat{\mathbf{J}}\rangle$ for the above state points in an arbitrary direction in space. Assuming that it lies in the first octant of the coordinate system, we perform a rotation of the coordinate system from $\{x, y, z\}$ to $\{x^\prime, y^\prime, z^\prime\}$, such that $\langle\hat{\mathbf{J}}\rangle$ points along the $z^\prime$ axis. The operators
$\{\hat{J}_{x^\prime}, \hat{J}_{y^\prime}, \hat{J}_{z^\prime}\}$ in the rotated frame $\{x^\prime, y^\prime, z^\prime\}$ are related to $\{\hat{J}_x, \hat{J}_y, \hat{J}_z\}$ in the unrotated frame $\{x, y, z\}$ as  

\begin{eqnarray}
\hat{J}_{x^\prime} &=& \hat{J}_x\cos\theta\cos\phi + \hat{J}_y
\cos\theta\sin\phi - \hat{J}_z\sin\theta\label{2.5d1}\\
\hat{J}_{y^\prime} &=& -\hat{J}_x\sin\phi + \hat{J}_y\cos\phi
\label{2.5d2}\\
\hat{J}_{z^\prime} &=& \hat{J}_x\sin\theta\cos\phi + \hat{J}_y
\sin\theta\sin\phi + \hat{J}_z\cos\theta,\label{2.5d3}
\end{eqnarray}
where,
\begin{eqnarray}
\cos\theta &=& \frac{\langle\hat{J}_z\rangle}
{|\langle\hat{\mathbf{J}}\rangle|}\label{2.5d4}\\
\cos\phi &=& \frac{\langle\hat{J}_x\rangle}{\sqrt{\langle\hat{J}_x\rangle^2 + \langle\hat{J}_y\rangle^2}}.
\label{2.5d5}
\end{eqnarray}
We can observe that for the above choice of $\cos\theta$ 
and $\cos\phi$, we have $\langle\hat{J}_{x^\prime}\rangle = 0$, $\langle\hat{J}_{y^\prime}\rangle = 0$, and the mean spin vector points along the $z^\prime$ axis.

Now, using Eqs. (\ref{2.5d1}) and (\ref{2.5d2})
the quantum fluctuations $\Delta{J_{x^\prime}^2}$ and
$\Delta{J_{y^\prime}^2}$ for this system are,
\begin{eqnarray}
&&\Delta{J_{x^\prime}^2} = \langle J_{x^\prime}^2 \rangle
- \langle J_{x^\prime} \rangle^2\nonumber\\ 
&=&\Delta{J_x}^2\cos^2\theta\cos^2\phi + \Delta{J_y}^2\cos^2\theta\sin^2\phi\nonumber \\
&+& \Delta{J_z}^2\sin^2\theta + \Big{(}\langle{J_x}{J_y} + J_y J_x\rangle - 2\langle J_x\rangle\langle J_y\rangle\Big{)}\nonumber\\
&\times&\cos^2\theta\sin\phi\cos\phi - \Big{(}\langle{J_x}{J_z} + J_z J_x\rangle - 2\langle J_x\rangle\langle J_z\rangle\Big{)}\nonumber\\
&\times&\sin\theta\cos\theta\cos\phi - \Big{(}\langle{J_y}{J_z} + J_z J_y\rangle - 2\langle J_y\rangle\langle J_z\rangle\Big{)}\nonumber\\
&\times&\sin\theta\cos\theta\sin\phi
\label{2.5d6}
\end{eqnarray}
and
\begin{eqnarray}
\Delta{J_{y^\prime}^2} &=& \langle J_{y^\prime}^2 \rangle
- \langle J_{y^\prime} \rangle^2\nonumber\\ 
&=&\Delta {J_x}^2\sin^2\phi + \Delta {J_y}^2\cos^2\phi - \Big{(}\langle{J_x}{J_y} + J_y J_x\rangle\nonumber\\
 &-& 2\langle J_x\rangle\langle J_y\rangle\Big{)}\sin\phi\cos\phi.
\label{2.5d7}
\end{eqnarray}
Now using Eqs. (\ref{1.3}), we have
\begin{eqnarray}
{\hat{J}_x}^2 &=& \sum_{i=1}^{N}\hat{J}_{i_{x}}^2 + \sum_{i=1}^{N}\sum_{{}^{l=1}_{l\ne i}}^{N}\hat{J}_{i_{x}}
\hat{J}_{l_{x}},\label{2.5d8}\\
{\hat{J}_y}^2 &=& \sum_{i=1}^{N}\hat{J}_{i_{y}}^2 + \sum_{i=1}^{N}\sum_{{}^{l=1}_{l\ne i}}^{N}\hat{J}_{i_{y}}
\hat{J}_{l_{y}},\label{2.5d9}\\
{\hat{J}_z}^2 &=& \sum_{i=1}^{N}\hat{J}_{i_{z}}^2 + \sum_{i=1}^{N}\sum_{{}^{l=1}_{l\ne i}}^{N}\hat{J}_{i_{z}}
\hat{J}_{l_{z}}\label{2.5d10}
\end{eqnarray}
and
\begin{eqnarray}
\langle\hat{J}_x\rangle^2 &=& \sum_{i=1}^{N}\langle
\hat{J}_{i_{x}}\rangle^2 + \sum_{i=1}^{N}\sum_{{}^{l=1}
_{l\ne i}}^{N}\langle\hat{J}_{i_{x}}\rangle
\langle\hat{J}_{l_{x}}\rangle\label{2.5d11}\\
\langle\hat{J}_y\rangle^2 &=& \sum_{i=1}^{N}\langle
\hat{J}_{i_{y}}\rangle^2 + \sum_{i=1}^{N}\sum_{{}^{l=1}
_{l\ne i}}^{N}\langle\hat{J}_{i_{y}}\rangle
\langle\hat{J}_{l_{y}}\rangle\label{2.5d12}\\
\langle\hat{J}_z\rangle^2 &=& \sum_{i=1}^{N}\langle
\hat{J}_{i_{z}}\rangle^2 + \sum_{i=1}^{N}\sum_{{}^{l=1}
_{l\ne i}}^{N}\langle\hat{J}_{i_{z}}\rangle
\langle\hat{J}_{l_{z}}\rangle.\label{2.5d13}
\end{eqnarray}

Using Eqs. (\ref{1.3}), we also have
\begin{eqnarray}
&&\langle{\hat{J}_x}{\hat{J}_y} + \hat{J}_y \hat{J}_x\rangle - 2\langle \hat{J}_x\rangle\langle \hat{J}_y\rangle
= \sum_{i=1}^{N}\Big{(}\langle\hat{J}_{i_{x}}\hat{J}_{i_{y}} + \hat{J}_{i_{y}} \hat{J}_{i_{x}}\rangle\nonumber\\
&-& 2\langle \hat{J}_{i_{x}}\rangle\langle \hat{J}_{i_{y}}\rangle\Big{)} + 2\sum_{i=1}^{N}\sum_{{}^{l=1}_{l\ne i}}^{N}\Big{(}\langle\hat{J}_{i_{x}}\hat{J}_{l_{y}}\rangle -
\langle\hat{J}_{i_{x}}\rangle\langle\hat{J}_{l_{y}}\rangle
\Big{)}
\label{2.5d14}
\end{eqnarray}
and
\begin{eqnarray}
&&\langle{\hat{J}_x}{\hat{J}_z} + \hat{J}_z \hat{J}_x\rangle - 2\langle \hat{J}_x\rangle\langle \hat{J}
_z\rangle
= \sum_{i=1}^{N}\Big{(}\langle\hat{J}_{i_{x}}\hat{J}_{i_{z}} + \hat{J}_{i_{z}} \hat{J}_{i_{x}}\rangle\nonumber\\
&-& 2\langle \hat{J}_{i_{x}}\rangle\langle \hat{J}_{i_{z}}\rangle\Big{)} + 2\sum_{i=1}^{N}\sum_{{}^{l=1}_{l\ne i}}^{N}\Big{(}\langle\hat{J}_{i_{x}}\hat{J}_{l_{z}}\rangle -
\langle\hat{J}_{i_{x}}\rangle\langle\hat{J}_{l_{z}}\rangle
\Big{)}
\label{2.5d15}
\end{eqnarray}
and
\begin{eqnarray}
&&\langle{\hat{J}_y}{\hat{J}_z} + \hat{J}_z \hat{J}_
y\rangle - 2\langle \hat{J}_y\rangle\langle \hat{J}_
z\rangle
= \sum_{i=1}^{N}\Big{(}\langle\hat{J}_{i_{y}}\hat{J}_{i_{z}} + \hat{J}_{i_{z}} \hat{J}_{i_{y}}\rangle\nonumber\\
&-& 2\langle \hat{J}_{i_{y}}\rangle\langle \hat{J}_{i_{z}}\rangle\Big{)} + 2\sum_{i=1}^{N}\sum_{{}^{l=1}_{l\ne i}}^{N}\Big{(}\langle\hat{J}_{i_{y}}\hat{J}_{l_{z}}\rangle -
\langle\hat{J}_{i_{y}}\rangle\langle\hat{J}_{l_{z}}\rangle
\Big{)}.
\label{2.5d16}
\end{eqnarray}
Now, using Eqs. (\ref{2.5d8}) - (\ref{2.5d16})
in Eq. (\ref{2.5d6}), we obtain
\begin{eqnarray}
&&\Delta J_{x^\prime}^2 = \Bigg{[}\sum_{i=1}^{N}
\langle\hat{J}_{i_{x}}^2\rangle + \sum_{i=1}^{N}\sum_{{}^{l=1}_{l\ne i}}^{N}\langle\hat{J}_{i_{x}}\hat{J}
_{l_{x}}\rangle - \Big{\{}\sum_{i=1}^{N}\langle\hat{J}_{i_{x}}\rangle^2 \nonumber\\
&+& \sum_{i=1}^{N}\sum_{{}^{l=1}_{l\ne i}}^{N}
\langle\hat{J}_{i_{x}}\rangle\langle\hat{J}_{l_{x}}
\rangle\Big{\}}\Bigg{]}\cos^2\theta\cos^2\phi + \Bigg{[}\sum_{i=1}^{N}
\langle\hat{J}_{i_{y}}^2\rangle\nonumber\\ 
&+& \sum_{i=1}^{N}\sum_{{}^{l=1}_{l\ne i}}^{N}\langle\hat{J}_{i_{y}}\hat{J}
_{l_{y}}\rangle - \Big{\{}\sum_{i=1}^{N}\langle\hat{J}_{i_{y}}\rangle^2\nonumber\\
&+& \sum_{i=1}^{N}\sum_{{}^{l=1}_{l\ne i}}^{N}
\langle\hat{J}_{i_{y}}\rangle\langle\hat{J}_{l_{y}}
\rangle\Big{\}}\Bigg{]}\cos^2\theta\sin^2\phi + \Bigg{[}\sum_{i=1}^{N}
\langle\hat{J}_{i_{z}}^2\rangle\nonumber\\ 
&+& \sum_{i=1}^{N}\sum_{{}^{l=1}_{l\ne i}}^{N}\langle\hat{J}_{i_{z}}\hat{J}
_{l_{z}}\rangle - \Big{\{}\sum_{i=1}^{N}\langle\hat{J}_{i_{z}}\rangle^2
+ \sum_{i=1}^{N}\sum_{{}^{l=1}_{l\ne i}}^{N}
\langle\hat{J}_{i_{z}}\rangle\langle\hat{J}_{l_{z}}
\rangle\Big{\}}\Bigg{]}\nonumber\\
&\times&\sin^2\theta
+\Bigg{[}\sum_{i=1}^{N}\Big{\{}\langle\hat{J}_{i_{x}}\hat{J}_{i_{y}} + \hat{J}_{i_{y}}\hat{J}_{i_{x}}\rangle
- 2\langle\hat{J}_{i_{x}}\rangle\langle\hat{J}_{i_{y}}\rangle  \Big{\}}\nonumber\\
&+& 2\sum_{i=1}^{N}\sum_{{}^{l=1}_{l\ne i}}^{N}\Big{\{}
\langle\hat{J}_{i_{x}}\hat{J}_{l_{y}}\rangle - \langle\hat{J}_
{i_{x}}\rangle\langle\hat{J}_{l_{y}}\rangle\Big{\}} \Bigg{]}\cos^2\theta\sin\phi\cos\phi\nonumber\\
&-& \Bigg{[}\sum_{i=1}^{N}\Big{\{}\langle\hat{J}_{i_{x}}\hat{J}_{i_{z}} + \hat{J}_{i_{z}}\hat{J}_{i_{x}}\rangle
- 2\langle\hat{J}_{i_{x}}\rangle\langle\hat{J}_{i_{z}}\rangle  \Big{\}}\nonumber\\
&+& 2\sum_{i=1}^{N}\sum_{{}^{l=1}_{l\ne i}}^{N}\Big{\{}
\langle\hat{J}_{i_{x}}\hat{J}_{l_{z}}\rangle - \langle\hat{J}_
{i_{x}}\rangle\langle\hat{J}_{l_{z}}\rangle\Big{\}} \Bigg{]}\sin\theta\cos\theta\cos\phi\nonumber\\
&-& \Bigg{[}\sum_{i=1}^{N}\Big{\{}\langle\hat{J}_{i_{y}}\hat{J}_{i_{z}} + \hat{J}_{i_{z}}\hat{J}_{i_{y}}\rangle
- 2\langle\hat{J}_{i_{y}}\rangle\langle\hat{J}_{i_{z}}\rangle  \Big{\}}\nonumber\\
&+& 2\sum_{i=1}^{N}\sum_{{}^{l=1}_{l\ne i}}^{N}\Big{\{}
\langle\hat{J}_{i_{y}}\hat{J}_{l_{z}}\rangle - \langle\hat{J}_
{i_{y}}\rangle\langle\hat{J}_{l_{z}}\rangle\Big{\}} \Bigg{]}\sin\theta\cos\theta\sin\phi.\nonumber
\label{2.5d17}\\    
\end{eqnarray}
The above expression can be rearranged and written as
\begin{eqnarray*}
&&\Delta J_{x^\prime}^2 = \sum_{i=1}^{N}\Bigg{[}\Delta
J_{i_{x}}^2\cos^2\theta\cos^2\phi + \Delta J_{i_{y}}^2\cos^2\theta\sin^2\phi  \nonumber\\
&+& \Delta J_{i_{z}}^2\sin^2\theta +\Big{(}\langle
\hat{J}_{i_{x}}\hat{J}_{i_{y}}+\hat{J}_{i_{y}}\hat{J}_
{i_{x}}\rangle - 2\langle\hat{J}_{i_{x}}\rangle\langle
\hat{J}_{i_{y}}\rangle\Big{)}\nonumber\\
&\times&\cos^2\theta\sin\phi\cos\phi - \Big{(}\langle
\hat{J}_{i_{x}}\hat{J}_{i_{z}}+\hat{J}_{i_{z}}\hat{J}_
{i_{x}}\rangle - 2\langle\hat{J}_{i_{x}}\rangle\langle
\hat{J}_{i_{z}}\rangle\Big{)}\nonumber\\ 
&\times&\sin\theta\cos\theta\cos\phi - \Big{(}\langle
\hat{J}_{i_{y}}\hat{J}_{i_{z}}+\hat{J}_{i_{z}}\hat{J}_
{i_{y}}\rangle - 2\langle\hat{J}_{i_{y}}\rangle\langle
\hat{J}_{i_{z}}\rangle\Big{)}
\end{eqnarray*}
\begin{eqnarray}
&\times&\sin\theta\cos\theta\sin\phi \Bigg{]} + 
\sum_{i=1}^{N} \sum_{{}^{l=1}_{l\ne i}}^{N}\Bigg{[}\Big{(}\langle\hat{J}_
{i_{x}}\hat{J}_{l_{x}}\rangle - \langle\hat{J}_
{i_{x}}\rangle\langle\hat{J}_{l_{x}}\rangle\Big{)}\nonumber\\
&\times&\cos^2\theta\cos^2\phi + \Big{(}\langle\hat{J}_
{i_{y}}\hat{J}_{l_{y}}\rangle - \langle\hat{J}_
{i_{y}}\rangle\langle\hat{J}_{l_{y}}\rangle\Big{)}\cos^2\theta\sin^2\phi\nonumber\\
&+& \Big{(}\langle\hat{J}_
{i_{z}}\hat{J}_{l_{z}}\rangle - \langle\hat{J}_
{i_{z}}\rangle\langle\hat{J}_{l_{z}}\rangle\Big{)}\sin^2\theta + 2\Big{(}\langle\hat{J}_
{i_{x}}\hat{J}_{l_{y}}\rangle\nonumber\\
&-& \langle\hat{J}_
{i_{x}}\rangle\langle\hat{J}_{l_{y}}\rangle\Big{)}\cos^2\theta\sin\phi\cos\phi - 2\Big{(}\langle\hat{J}_
{i_{x}}\hat{J}_{l_{z}}\rangle\nonumber\\
 &-& \langle\hat{J}_
{i_{x}}\rangle\langle\hat{J}_{l_{z}}\rangle\Big{)}\sin\theta\cos\theta\cos\phi - 2\Big{(}\langle\hat{J}_
{i_{y}}\hat{J}_{l_{z}}\rangle - \langle\hat{J}_
{i_{y}}\rangle\langle\hat{J}_{l_{z}}\rangle\Big{)}\nonumber\\
&\times& \sin\theta\cos\theta\sin\phi\Bigg{]}.
\label{2.5d18}
\end{eqnarray}
If, we now compare the term under the single summation symbol with Eq. (\ref{2.5d6}), we find that it is
$\Delta J_{i_{x^\prime}}^2$.
Therefore, Eq. (\ref{2.5d18}) can be expressed as,
\begin{equation}
\Delta J_{x^\prime}^2 = \sum_{i=1}^{N}\Delta J_{i_{x^\prime}}^2 + CORRX,
\label{2.5d19}
\end{equation} 
where $CORRX$ is the term with the double summation symbol in Eq. (\ref{2.5d18}). We notice that $CORRX$ is solely made up of the quantum correlations among the 
$N$ two level atoms.   
Thus, the quantum fluctuation $\Delta J_{x^\prime}^2$ for the composite state of $N$ two-level atoms has been expressed as the algebraic sum of the quantum fluctuations 
$\Delta J_{i_{x^\prime}}^2$ of the $N$ individual atoms 
and the quantum correlation term $CORRX$ among the $N$ atoms. We now present a similar expression for 
$\Delta J_{y^\prime}^2$. Using Eqs. (\ref{2.5d8}), (\ref{2.5d9}), (\ref{2.5d11}), (\ref{2.5d12}), and (\ref{2.5d14}) in Eq. (\ref{2.5d7}), we obtain
\begin{eqnarray}
&&\Delta J_{y^\prime}^2 = \Bigg{[}\sum_{i=1}^{N}
\langle\hat{J}_{i_{x}}^2\rangle + \sum_{i=1}^{N}\sum_{{}^{l=1}_{l\ne i}}^{N}\langle\hat{J}_{i_{x}}\hat{J}
_{l_{x}}\rangle - \Big{\{}\sum_{i=1}^{N}\langle\hat{J}_{i_{x}}\rangle^2 \nonumber\\
&+& \sum_{i=1}^{N}\sum_{{}^{l=1}_{l\ne i}}^{N}
\langle\hat{J}_{i_{x}}\rangle\langle\hat{J}_{l_{x}}
\rangle\Big{\}}\Bigg{]}\sin^2\phi + \Bigg{[}\sum_{i=1}^{N}
\langle\hat{J}_{i_{y}}^2\rangle\nonumber\\ 
&+& \sum_{i=1}^{N}\sum_{{}^{l=1}_{l\ne i}}^{N}\langle\hat{J}_{i_{y}}\hat{J}
_{l_{y}}\rangle - \Big{\{}\sum_{i=1}^{N}\langle\hat{J}_{i_{y}}\rangle^2
+ \sum_{i=1}^{N}\sum_{{}^{l=1}_{l\ne i}}^{N}
\langle\hat{J}_{i_{y}}\rangle\langle\hat{J}_{l_{y}}
\rangle\Big{\}}\Bigg{]}\nonumber\\
&\times&\cos^2\phi - \Bigg{[}\sum_{i=1}^{N}\Big{\{}\langle\hat{J}_{i_{x}}\hat{J}_{i_{y}} + \hat{J}_{i_{y}}\hat{J}_{i_{x}}\rangle
- 2\langle\hat{J}_{i_{x}}\rangle\langle\hat{J}_{i_{y}}\rangle  \Big{\}}\nonumber\\
&+& 2\sum_{i=1}^{N}\sum_{{}^{l=1}_{l\ne i}}^{N}\Big{\{}
\langle\hat{J}_{i_{x}}\hat{J}_{l_{y}}\rangle - \langle\hat{J}_
{i_{x}}\rangle\langle\hat{J}_{l_{y}}\rangle\Big{\}}\Bigg{]}\sin\phi\cos\phi.
\label{2.5d20}
\end{eqnarray}
The above expression can be rearranged as
\begin{eqnarray*}
&&\Delta J_{y^\prime}^2 = \sum_{i=1}^{N}\Bigg{[}\Delta
J_{i_{x}}^2\sin^2\phi + \Delta J_{i_{y}}^2\cos^2\phi  \nonumber\\
&-& \Big{(}\langle
\hat{J}_{i_{x}}\hat{J}_{i_{y}}+\hat{J}_{i_{y}}\hat{J}_
{i_{x}}\rangle - 2\langle\hat{J}_{i_{x}}\rangle\langle
\hat{J}_{i_{y}}\rangle\Big{)}
\sin\phi\cos\phi\nonumber\Bigg{]}
\end{eqnarray*}
\begin{eqnarray} 
&+& \sum_{i=1}^{N} \sum_{{}^{l=1}_{l\ne i}}^{N}\Bigg{[}\Big{(}\langle\hat{J}_
{i_{x}}\hat{J}_{l_{x}}\rangle - \langle\hat{J}_
{i_{x}}\rangle\langle\hat{J}_{l_{x}}\rangle\Big{)}
\sin^2\phi\nonumber\\
&+& \Big{(}\langle\hat{J}_
{i_{y}}\hat{J}_{l_{y}}\rangle - \langle\hat{J}_
{i_{y}}\rangle\langle\hat{J}_{l_{y}}\rangle\Big{)}
\cos^2\phi
+ 2\Big{(}\langle\hat{J}_
{i_{x}}\hat{J}_{l_{y}}\rangle\nonumber\\
&-& \langle\hat{J}_
{i_{x}}\rangle\langle\hat{J}_{l_{y}}\rangle\Big{)}\sin\phi\cos\phi\Bigg{]}.
\label{2.5d21}
\end{eqnarray}
Comparing the term under the single summation symbol in
Eq. (\ref{2.5d21}) with the right hand side of Eq. (\ref{2.5d7}), we notice that it is 
$\Delta J_{i_{y^\prime}}^2$. Thus, Eq. (\ref{2.5d21}) can be written as
\begin{equation}
\Delta J_{y^\prime}^2 = \sum_{i=1}^{N}\Delta J_{i_{y^\prime}}^2 + CORRY,
\label{2.5d22}
\end{equation}
where, $CORRY$ represents the term with the double summation symbol in Eq. (\ref{2.5d21}). We notice that 
$CORRY$ is solely made up of the quantum correlations among the $N$ two-level atoms. 
Thus, the quantum fluctuation $\Delta J_{y^\prime}^2$ for the composite state of $N$ two-level atoms has been expressed as the algebraic sum of the quantum fluctuations $\Delta J_{i_{y^\prime}}^2$ of the $N$ individual constituent atoms and the quantum correlation term among them.

Now if the quantum state $|\Psi\rangle$ is an unentangled state, that is
\begin{equation}
|\Psi\rangle = |\Psi_1\rangle \otimes |\Psi_2\rangle \otimes |\Psi_3\rangle......\otimes |\Psi_N\rangle,
\label{2.5d23}
\end{equation}
where $|\Psi_1\rangle$, $|\Psi_2\rangle$, ...
$|\Psi_N\rangle$ are the individual atomic state vectors,
then, we have for $i \ne l$,
\begin{eqnarray}
\langle\hat{J}_{i_{x}}\hat{J}_{l_{x}}\rangle &=& \langle
\hat{J}_{i_{x}}\rangle\langle\hat{J}_{l_{x}}\rangle,~~~~
\langle\hat{J}_{i_{y}}\hat{J}_{l_{y}}\rangle = \langle
\hat{J}_{i_{y}}\rangle\langle\hat{J}_{l_{y}}
\rangle,\nonumber\\
\langle\hat{J}_{i_{z}}\hat{J}_{l_{z}}\rangle &=& \langle
\hat{J}_{i_{z}}\rangle\langle\hat{J}_{l_{z}}\rangle,~~~~
\langle\hat{J}_{i_{x}}\hat{J}_{l_{y}}\rangle = \langle
\hat{J}_{i_{x}}\rangle\langle\hat{J}_{l_{y}}\rangle,
\nonumber\\
\langle\hat{J}_{i_{x}}\hat{J}_{l_{z}}\rangle &=& \langle
\hat{J}_{i_{x}}\rangle\langle\hat{J}_{l_{z}}\rangle,~~~~
\langle\hat{J}_{i_{y}}\hat{J}_{l_{z}}\rangle = \langle
\hat{J}_{i_{y}}\rangle\langle\hat{J}_{l_{z}}\rangle.
\label{2.5d24}
\end{eqnarray}
In this case the terms under the double summation symbols in Eqs. (\ref{2.5d18}) and (\ref{2.5d21}) are zero, implying that
\begin{equation}
CORRX = CORRY = 0.
\label{2.5d25}
\end{equation} 
Thus, for an unentangled state we have $CORRX = CORRY = 0$.

Now, we can calculate and find that the quantum fluctuations 
$\Delta{J_{i_{x^\prime}}}^2$ and 
$\Delta{J_{i_{y^\prime}}}^2$ of the individual constituent atoms have value $\frac{1}{4}$, that is

\begin{eqnarray}
\Delta{J_{1_{x^\prime}}}^2 &=& \Delta{J_{2_{x^\prime}}}^2
=...\Delta{J_{N_{x^\prime}}}^2 = \frac{1}{4},
\label{2.8}\\
\Delta{J_{1_{y^\prime}}}^2 &=& \Delta{J_{2_{y^\prime}}}^2
=...\Delta{J_{N_{y^\prime}}}^2 = \frac{1}{4}.
\label{2.9}
\end{eqnarray} 
Therefore, using Eqs. (\ref{2.8}) and (\ref{2.9}) in Eqs. (\ref{2.5d19}) and (\ref{2.5d22}), we obtain
\begin{eqnarray}
\Delta {J_{x^\prime}}^2  &=& \frac{N}{4} + CORRX,
\label{2.10a1}\\
\Delta {J_{y^\prime}}^2 &=&  \frac{N}{4} + CORRY.
\label{2.10b1}
\end{eqnarray}
Now, it has been discussed that \cite{Kitagawa}, $N/4$ is the quantum fluctuation in $\hat{J}_{x^\prime}$ and $\hat{J}_{y^\prime}$ for a system of $N$
two-level atoms in a coherent spin state (CSS), which is a completely separable state, that is, an unentangled state. 
So, we can write Eqs. (\ref{2.10a1}) and (\ref{2.10b1}) as
\begin{eqnarray}
\Delta {J_{x^\prime}}^2 - \Delta {J_{x^\prime}}^2|_{un-ent.} &=& CORRX~,\label{2.10a}\\
\Delta {J_{y^\prime}}^2 - \Delta {J_{y^\prime}}^2|_{un-ent.} &=& CORRY.
\label{2.10b}
\end{eqnarray}

Therefore, $CORRX$ and 
$CORRY$ are the measures of the deviations of the quantum fluctuations in $\hat{J}_{x^\prime}$ and 
$\hat{J}_{y^\prime}$ respectively, of the state 
$|\Psi\rangle$ from those of an unentangled state.

If for some quantum state of the $N$ two-level atoms 
 we have $CORRX < 0$, then $\Delta {J_{x^\prime}}^2 < N/4$, and we say that the corresponding quantum state has spin squeezing in the $x^\prime$ quadrature. In that case we must have $CORRY > 0$, as to restore the Heisenberg's uncertainty principle (Eq. (\ref{2.5b11})). Similarly, when $CORRY < 0$, making $CORRX > 0$, we say that there is spin squeezing in the $y^\prime$ quadrature.

The non-zero value of either $CORRX$ or $CORRY$ or both implies the presence of quantum correlations among the atoms, because in that case the conditions in Eq. (\ref{2.5d24}) are not all satisfied and the terms under the 
double summation symbols in Eqs. (\ref{2.5d18}) and (\ref{2.5d21}) are non-zero.
 Now, it may happen that for some quantum state of $N$ two-level atoms, both $CORRX$ and $CORRY$ are greater than zero, implying no spin squeezing at all, but since
$CORRX$ and $CORRY$ are non-zero, there is  quantum correlations among the atoms and we have entanglement. Thus, the presence of entanglement cannot be detected always by spin squeezing and we show this for a real physical system in the next section.

We notice from Eqs. (\ref{2.5d18}), (\ref{2.5d19}), 
(\ref{2.5d21}), and (\ref{2.5d22}), that $CORRX$ and 
$CORRY$ are the extracts from the composite quantum fluctuations $\Delta J_{x^\prime}^2$ and 
$\Delta J_{y^\prime}^2$ of the $N$ two-level atoms, representing only the quantum correlations among the 
$N$ atoms. We also observe from the above mentioned equations that $CORRX$ and $CORRY$ are made up of all possible bipartite quantum correlations among the $N$ atoms.
Therefore, we use these correlation terms, $CORRX$ and $CORRY$,
to construct the multiparticle quantum entanglement parameter. Now, since, $CORRX$ and $CORRY$ may be positive or negative, we construct the multiparticle quantum entanglement parameter $E$ as
\begin{equation}
E = \frac{1}{2}\Big{[}(CORRX)^2 + (CORRY)^2\Big{]}.
\label{2.11}
\end{equation}   
Thus, $E$ is the mean squared deviation of the quantum fluctuations in $\hat{J}_{x^\prime}$ and 
$\hat{J}_{y^\prime}$ of an arbitrary state from those of an unentangled state. Now, using Eqs. (\ref{2.10a1}) and (\ref{2.10b1}), we can express $E$ as,
\begin{eqnarray}
E &=& \frac{1}{2}\Big{[}\Delta {J_{x^\prime}}^2\Big{(}\Delta {J_{x^\prime}}^2 -\frac{N}{2}\Big{)} + \Delta {J_{y^\prime}}^2\Big{(}\Delta {J_{y^\prime}}^2-\frac{N}{2}\Big{)}\nonumber\\ 
&+& \frac{N^2}{8}\Big{]}.
\label{2.12}
\end{eqnarray}
Now, if $\xi_{R_{x}}$ and $\xi_{R_{y}}$ are the spectroscopic squeezing parameters used in Ramsey spectroscopy \cite{Wineland}, given as
\begin{equation}
\xi_{R_{x}} = \frac{\sqrt{2j}}{|\langle{\bf \hat{J}}\rangle|}\Delta J_{x^\prime}~,~~~~ \xi_{R_{y}} = \frac{\sqrt{2j}}{|\langle{\bf \hat{J}}\rangle|}
\Delta J_{y^\prime},
\label{2.13}
\end{equation}
then,
\begin{eqnarray}
E &=& \frac{1}{2}\Bigg[ \frac{\xi_{R_x}^2|\langle\hat{\mathbf J}\rangle|^2}{2j} 
\bigg( \frac{\xi_{R_x}^2|\langle\hat{\mathbf J}\rangle|^2}{2j} - 
\frac{N}{2} \bigg)\nonumber\\
&& + ~~ 
\frac{\xi_{R_y}^2|\langle\hat{\mathbf J}\rangle|^2}{2j} 
\bigg(\frac{\xi_{R_y}^2|\langle\hat{\mathbf J}\rangle|^2}{2j} - 
\frac{N}{2} \bigg) + 
\frac{N^2}{8}\Bigg].
\label{2.14}
\end{eqnarray}
Thus, the quantum entanglement parameter $E$ is connected 
to the experimentally measurable quantities. 

Now, if for a quantum state of $N$ two-level atoms, $E = 0$, then, both $CORRX$ and $CORRY$ are zero, that is all the conditions in Eqs. (\ref{2.5d24}) are satisfied and the corresponding quantum state is completely separable, and we have no entanglement. If $E > 0$, then either $CORRX$ or $CORRY$ or both are non zero, and the corresponding quantum state is an entangled state. Now, the numerical value of $E$ can be a measure of the amount of quantum entanglement present among the $N$ two-level atoms. Thus, by measuring $E$ with the help of Eqs. (\ref{2.12}) or (\ref{2.14}) for a quantum state of $N$ two-level atoms we can measure the amount of quantum entanglement present in that system. In the following section we apply this idea to measure the amount of quantum entanglement present in a real physical system, which is of considerable interest in Quantum Optics.

\section{III. Quantum entanglement of a system of $N$ two-level atoms interacting with the squeezed vacuum state of the radiation field}

We consider a system of $N$ two-level atoms in interaction with the squeezed vacuum state of the electromagnetic field. 
The composite quantum state vector for such system is given as \cite{Agarwal1}, \cite{Agarwal},
\begin{equation}
|\Psi_m\rangle = A_m e^{\xi\hat{J}_z} e^{-i\frac{\pi}{2}\hat{J}_y}|j,m\rangle,
\label{3.1}
\end{equation} 
where $A_m$ is the normalization constant, $\xi$ is the radiation field parameter and $|j,m\rangle$ is the Wigner state. 
Inserting the identity operator
\begin{equation}
\sum_{m^\prime = -j}^{j} |j,m^\prime\rangle\langle j,m^\prime| = I
\label{3.1a1}
\end{equation} 
after the operator $e^{\xi\hat{J}_z}$ in Eq. (\ref{3.1}), we get
\begin{eqnarray}
|\Psi_m\rangle &=& A_m e^{\xi\hat{J}_z}\sum_{m^\prime = -j}^{j} |j,m^\prime\rangle\langle j,m^\prime|e^{-i\frac{\pi}{2}\hat{J}_y}|j,m\rangle\nonumber\\
&=& A_m e^{\xi\hat{J}_z} \sum_{m^\prime = -j}^{j}d^{j}_{m^\prime m}(\pi/2)|j,m^\prime\rangle\nonumber\\
&=& A_m \sum_{m^\prime = -j}^{j} e^{\xi m^\prime}d^{j}_{m^\prime m}(\pi/2)|j,m^\prime\rangle,
\label{3.1a2}
\end{eqnarray}
where $d^{j}_{m^\prime m}(\pi/2)$ is the reduced Wigner
$d$-matrix element \cite{Sakurai} given as,
\begin{eqnarray}
&&d_{m^\prime m}^j (\beta) =\langle j,m^\prime\vert 
\hat{R}_{y}(\beta)
\vert j,m\rangle =\langle j,m^\prime\vert e^{-i\beta J_y}
\vert j,m\rangle \nonumber\\\nonumber\\  
 &=& (-1)^{m^\prime -m}\sqrt{(j+m)!(j-m)!(j+m^\prime )!
(j-m^\prime )!}\nonumber \\\nonumber\\ 
&\times & \sum_{k}\frac{(-1)^k (\cos\frac{\beta}{2})^{2j-2k-m^\prime +m}
(\sin\frac{\beta}{2})^{2k+m^\prime -m}}{k!(j-m^\prime -k)!(j+m-k)!(m^\prime -m+k)!}.
\label{3.1a3}
\end{eqnarray}

The normalization constant $A_m$ is given by
\begin{equation}
A_m^{-2} = \sum_{m^\prime = -j}^{j} e^{2\xi m^\prime}
d^j_{m^\prime m}(\pi/2)d^j_{m^\prime m}(\pi/2).
\label{3.1a4}
\end{equation}
Using the addition theorem and symmetry properties of Wigner rotation matrices \cite{Varshalovich}, we can obtain
\begin{equation}
A_m^{-2} = d^j_{mm}(2i\xi) \equiv \Delta,
\label{3.1a5}
\end{equation}
where $d^j_{mm}(2i\xi) \equiv \Delta$, is the reduced Wigner $d$-matrix element for imaginary angle
$2i\xi$, and is given as
\begin{equation}
\Delta = (j+m)!(j-m)!
\sum_{k}\frac{(\cosh\xi)^{2j}(\tanh\xi)^{2k}}{(k!)^2(j-m-k)!(j+m-k)!},
\label{3.4}
\end{equation}

In Refs. \cite{Agarwal} and \cite{Ram1}, the spin squeezing properties of the state $|\Psi_m\rangle$ have been studied. But, quantum entanglement does not have any relationship with spin squeezing, which we show below.
In this paper, we study the quantum entanglement property of this state.

Now, to calculate and quantify the amount of quantum 
entanglement among the $N$ two-level atoms in this state, we need to calculate the quantum entanglement parameter $E$ for this state. This in turn, requires the calculation of the quantum fluctuations $\Delta J_{x^\prime}^2$ and 
$\Delta J_{y^\prime}^2$ for this state in a coordinate frame in which the mean pseudo-spin vector 
$\langle{\bf \hat{J}}\rangle$ points along the $z^\prime$-axis. Therefore, we now calculate 
$\Delta J_{x^\prime}^2$ and $\Delta J_{y^\prime}^2$ for the state $|\Psi_m\rangle$.
Now, $|\Psi_m\rangle$ is an eigenvector of a non-Hermitian operator $\hat{\Lambda}$ with eigenvalue 
$m$, \cite{Agarwal1}, \cite{Agarwal}  as
\begin{equation}
 \hat{\Lambda} |\Psi_{m}\rangle = ( \hat{J}_{x}\cosh\xi + 
i\hat{J}_{y}\sinh\xi)|\Psi_{m}\rangle
= m |\Psi_{m}\rangle.
\label{2.3.1}
\end{equation}
 The dual of the above is written as 
\begin{equation}
 \langle\Psi_{m}| \hat{\Lambda}^{\dagger} = \langle\Psi_{m}| (\hat{J}_{x}\cosh\xi - 
i\hat{J}_{y}\sinh\xi) = \langle\Psi_{m}|m.
\label{2.3.2} 
\end{equation}
Taking the scalar product of both sides of Eq. 
$(\ref{2.3.1})$ by
 $ \langle\Psi_{m}|$
and equating the real and imaginary parts we get,
\begin{equation}
\langle\Psi_{m}| \hat{J}_{x} |\Psi_{m}\rangle = 
\frac{m}{\cosh\xi}  
\label{2.3.3}
\end{equation}
and
\begin{equation}
\langle\Psi_{m}| \hat{J}_{y} |\Psi_{m}\rangle = 0. 
\label{2.3.4}
\end{equation}
Taking the scalar product of Eq. $(\ref{2.3.1})$ with it's dual i.e.
Eq. $(\ref{2.3.2})$ viz.
\begin{equation}
\langle\Psi_{m}| \hat{\Lambda}^{\dagger}\hat{\Lambda} 
|\Psi_{m}\rangle = m^2,
\label{2.3.4new1}
\end{equation} 
we obtain
\begin{eqnarray}
&&\cosh^{2}\xi\langle\Psi_{m}|{\hat{J}_{x}}^2
|\Psi_{m}\rangle + \sinh^{2}\xi
\langle\Psi_{m}|{\hat{J}_{y}}^2|\Psi_{m}\rangle\nonumber\\
&+& i\sinh\xi\cosh\xi
\langle\Psi_{m}| [\hat{J}_{x},\hat{J}_{y}] |\Psi_{m}
\rangle = m^2
\label{3.1a6} 
\end{eqnarray}
or,
\begin{eqnarray}
&&\cosh^{2}\xi\langle\Psi_{m}|{\hat{J}_{x}}^2|\Psi_{m}\rangle + \sinh^{2}\xi
\langle\Psi_{m}|{\hat{J}_{y}}^2|\Psi_{m}\rangle\nonumber\\
&-& \sinh\xi\cosh\xi\langle\Psi_{m}|
\hat{J}_{z}|\Psi_{m}\rangle = m^2.
\label{2.3.5}
\end{eqnarray}

Operating with $\hat{\Lambda}$ on both sides of 
Eq. $(\ref{2.3.1})$ and then taking the 
scalar product with $\langle\Psi_{m}|$ we get 
\begin{equation}
\langle\Psi_{m}| {\hat{\Lambda}}^2 |\Psi_{m}\rangle = m^2.
\label{2.3.6new1} 
\end{equation}
More explicitly it is
\begin{eqnarray}
&&\cosh^{2}\xi\langle\Psi_{m}|{\hat{J}_{x}}^2|
\Psi_{m}\rangle - \sinh^{2}\xi
\langle\Psi_{m}|{\hat{J}_{y}}^2|\Psi_{m}\rangle\nonumber\\
&+& i\sinh\xi\cosh\xi\langle\Psi_{m}|
\hat{J}_{x}\hat{J}_{y} + \hat{J}_{y}\hat{J}_{x}|\Psi_{m}\rangle = m^2. 
\end{eqnarray}
Equating the real and imaginary parts we obtain
\begin{equation}
\cosh^{2}\xi\langle\Psi_{m}|{\hat{J}_{x}}^2|\Psi_{m}\rangle - \sinh^{2}\xi
\langle\Psi_{m}|{\hat{J}_{y}}^2|\Psi_{m}\rangle = m^2 
\label{2.3.6}
\end{equation}
and
\begin{equation}
\langle\Psi_{m}| \hat{J}_{x}\hat{J}_{y} + 
\hat{J}_{y}\hat{J}_{x} |\Psi_{m}\rangle = 0. 
\label{2.3.7}
\end{equation}
 We now operate both sides of Eq. $(\ref{2.3.1})$ by 
$\hat{J}_{z}$ from left and then 
take the scalar product with $\langle\Psi_{m}|$, and obtain
\begin{equation}
\langle\Psi_{m}| \hat{J}_{z} (\hat{J}_{x}\cosh\xi + 
i\hat{J}_{y}\sinh\xi) |\Psi_{m}\rangle = 
m \langle\Psi_{m}| \hat{J}_{z} |\Psi_{m}\rangle. 
\label{2.3.8}
\end{equation}

Taking complex conjugate of both sides, we obtain
\begin{equation}
 \langle\Psi_{m}| (\hat{J}_{x}\cosh\xi - i\hat{J}_{y}\sinh\xi) \hat{J}_{z} |\Psi_{m}\rangle =
m \langle\Psi_{m}| \hat{J}_{z} |\Psi_{m}\rangle. 
\label{2.3.9}
\end{equation}

Adding Eqs. $(\ref{2.3.8})$ and $(\ref{2.3.9})$, we get
\begin{eqnarray}
&&\cosh\xi \langle\Psi_{m}| \hat{J}_{z}\hat{J}_{x} + 
\hat{J}_{x}\hat{J}_{z} |\Psi_{m}\rangle \nonumber\\
&+&i\sinh\xi\langle\Psi_{m}| [\hat{J}_{z},\hat{J}_{y}] |\Psi_{m}\rangle = 2m \langle\Psi_{m}|
\hat{J}_{z} |\Psi_{m}\rangle
\label{3.1a7}
\end{eqnarray} 
or,
\begin{eqnarray}
&&\langle\Psi_{m}| \hat{J}_{x}\hat{J}_{z} + \hat{J}_{z}
\hat{J}_{x} |\Psi_{m}\rangle = 
\frac{2m}{\cosh\xi} \langle\Psi_{m}| \hat{J}_{z} |\Psi_{m}\rangle\nonumber\\
&-& \frac{\sinh\xi}
{\cosh\xi} \langle\Psi_{m}| \hat{J}_{x} |\Psi_{m}\rangle.
\label{3.1a8}
\end{eqnarray}

Using Eq. $(\ref{2.3.3})$ we have
\begin{eqnarray}
\langle\Psi_{m}| \hat{J}_{x}\hat{J}_{z} + \hat{J}_{z}
\hat{J}_{x} |\Psi_{m}\rangle &=&
\frac{2m}{\cosh\xi} \langle\Psi_{m}| \hat{J}_{z} |\Psi_{m}\rangle\nonumber\\
&-& m \frac{\sinh\xi}{\cosh^{2}\xi}.
\label{2.3.10}
\end{eqnarray}

Subtracting Eq. $(\ref{2.3.9})$ from Eq. $(\ref{2.3.8})$ we have
\begin{eqnarray}
&&\langle\Psi_{m}| [\hat{J}_{z},\hat{J}_{x}] |\Psi_{m}\rangle \cosh\xi \nonumber\\
&+& i\sinh\xi\langle
\Psi_{m}| \hat{J}_{y}\hat{J}_{z} + \hat{J}_{z}\hat{J}_{y} 
|\Psi_{m}\rangle
 = 0
\label{3.1a9} 
\end{eqnarray}
or,
$$  \langle\Psi_{m}| \hat{J}_{y} |\Psi_{m}\rangle \cosh\xi + \sinh\xi\langle\Psi_{m}|
\hat{J}_{y}\hat{J}_{z} + \hat{J}_{z}\hat{J}_{y} |\Psi_{m}\rangle = 0. $$ 

Using Eq. $(\ref{2.3.4})$ we obtain
\begin{equation}
\langle\Psi_{m}|\hat{J}_{y}\hat{J}_{z} + 
\hat{J}_{z}\hat{J}_{y} |\Psi_{m}\rangle = 0.
\label{2.3.11}
\end{equation}
To calculate $\langle\Psi_{m}| \hat{J}_{z} |\Psi_{m}\rangle$ we proceed as below.
\begin{eqnarray}
\hat{J}_{z}|\Psi_{m}\rangle &=& \hat{J}_{z} A_{m}
\sum_{m^{\prime}=-j}^{+j} 
d^{j}_{m^{\prime}m}(\frac{\pi}{2}) e^{\xi{m^{\prime}}}|j,m^{\prime}\rangle\nonumber\\
&=& A_{m} \sum_{m^{\prime}=-j}^{+j}
d^{j}_{m^{\prime}m}(\frac{\pi}{2}) e^{\xi{m^{\prime}}} m^{\prime}|j,m^{\prime}
\rangle.
\label{3.1a10}
\end{eqnarray}

Taking scalar product by $\langle\Psi_{m}|$ we obtain
\begin{eqnarray}
&&\langle\Psi_{m}| \hat{J}_{z} |\Psi_{m}\rangle = |A_{m}|^2 \sum_{m^{\prime}=-j}^{+j}
\sum_{m^{\prime\prime}=-j}^{+j} d^{j}_{m^{\prime\prime}m}(\frac{\pi}{2})\nonumber\\
&\times&d^{j}_{m^{\prime}m}(\frac{\pi}{2}) e^{\xi(m^{\prime}+m^{\prime\prime})} 
m^{\prime}\langle{j,m^{\prime\prime}}|j,m^{\prime}\rangle
\nonumber\\  
&=& |A_{m}|^2 \sum_{m^{\prime}=-j}^{+j} d^{j}_{m^{\prime}m}
(\frac{\pi}{2})d^{j}_{m^{\prime}m}(\frac{\pi}{2})e^{2\xi{m^{\prime}}}m^{\prime}\nonumber\\
&=& |A_{m}|^2 \frac{1}{2}\frac{d}{d\xi} \sum_{m^{\prime}=-j}^{+j} 
d^{j}_{mm^{\prime}} (-\frac{\pi}{2})d^{j}_{m^{\prime}m}(\frac{\pi}{2})
e^{2\xi{m^{\prime}}}\nonumber\\
&=& |A_{m}|^2 \frac{1}{2}\frac{d}{d\xi}d^{j}_{mm}(2i\xi).
\label{2.3.12a1}
\end{eqnarray} 
Using Eq. $(\ref{3.1a5})$ we obtain
\begin{equation}
\langle\Psi_{m}| \hat{J}_{z} |\Psi_{m}\rangle = \frac{1}{2\Delta}\frac{d\Delta}
{d\xi}. 
\label{2.3.12}
\end{equation}
 
With this expression of $\langle{\hat{J}_{z}}\rangle$, 
Eq. $(\ref{2.3.10})$ becomes
\begin{equation}
\langle\Psi_{m}|\hat{J}_{x}\hat{J}_{z} + 
\hat{J}_{z}\hat{J}_{x}|\Psi_{m}\rangle = \frac{m}{\cosh\xi}
\frac{1}{\Delta}\frac{d\Delta}{d\xi} - m\frac{\sinh\xi}
{\cosh^{2}\xi}.
\label{2.3.13}
\end{equation}
In the same manner it can be shown that
\begin{equation}
\langle\Psi_{m}| \hat{J}_{z}^2 |\Psi_{m}\rangle = |A_{m}|^2\frac{d^2}{d\xi^2}
[d^{j}_{mm}(2i\xi)] = \frac{1}{4\Delta}\frac{d^2\Delta}{d\xi^2}. 
\label{2.3.14}
\end{equation}

Adding Eqs. $(\ref{2.3.5})$ and $(\ref{2.3.6})$ and using 
Eq. $(\ref{2.3.12})$ we get
\begin{equation}
\langle\Psi_{m}|\hat{J}_{x}^{2}|\Psi_{m}\rangle = \frac{m^2}{{\cosh^2}\xi} +
\frac{1}{4\Delta}\tanh\xi\frac{d\Delta}{d\xi}. 
\label{2.3.15}
\end{equation}

Subtracting Eq. $(\ref{2.3.6})$ from Eq. $(\ref{2.3.5})$ and using 
Eq. $(\ref{2.3.12})$ we obtain
\begin{equation}
\langle\Psi_{m}|{\hat{J}_{y}}^2|\Psi_{m}\rangle = \frac{1}{4\Delta}\coth\xi\frac
{d\Delta}{d\xi}. 
\label{2.3.16}
\end{equation}

The quantity $\frac{d\Delta}{d\xi}$ is
expressed in suitable form. Differentiating once the expression of $\Delta$ as
given in Eq. $(\ref{3.4})$, we obtain
\begin{equation}
\frac{d\Delta}{d\xi} = \tanh\xi \Gamma
\label{2.3.17}
\end{equation}
 with
\begin{equation}
\Gamma = 2j\Delta + 2\frac{\eta}{{\cosh^2}\xi} 
\label{2.3.18}
\end{equation}
where $\eta$ is given as
\begin{eqnarray}
\eta &=& (\cosh\xi)^{2j}(j+m)!(j-m)!\nonumber\\
&\times&\sum_{k}^{}\frac{(\tanh\xi)^{2k}}{k!(k+1)!
(j+m-1-k)!(j-m-1-k)!}.\nonumber\\ 
\label{2.3.19}
\end{eqnarray}
To express $\frac{d^2\Delta}{d\xi^2}$ in suitable form we use the differential
equation satisfied by the rotation matrix element 
$D^{j}_{m^{\prime}m}(\alpha, \beta, \gamma)$ familiar from
the quantum mechanics of a symmetric top \cite{Varshalovich}. The rotation matrix element $D^{j}_{m^{\prime}m}(\alpha, \beta, \gamma)$ is defined as
\begin{equation}
R(\alpha, \beta, \gamma)|j,m\rangle = \sum_{m^\prime}^{} D^j_{m^\prime m}(\alpha, \beta, \gamma)|j, m^\prime\rangle, 
\end{equation} 
where $R(\alpha, \beta, \gamma)$ is the rotation matrix and $\alpha, \beta, \gamma$ are the Euler angles. Now, the differential equation satisfied by
$D^{j}_{m^{\prime}m}(\alpha, \beta, \gamma)$ is
\begin{eqnarray}
&\Bigg{[}&-\frac{1}{\sin\beta}\frac{\partial}{\partial \beta}\sin\beta\frac{\partial}{\partial\beta} + \frac{m^2 - 2mm^\prime\cos\beta + {m^\prime}^2}{\sin^2
\beta}\Bigg{]}\nonumber\\
&\times& D^{j}_{m{m^\prime}}(\alpha, \beta, \gamma) = j(j+1) D^{j}_{m{m^\prime}}(\alpha, \beta, \gamma).
\label{3.1a12}
\end{eqnarray}
Now, by putting $m^\prime = m$ and $\alpha = \gamma = 0$ we get
\begin{eqnarray}
\frac{d^2}{d\beta^2}d^j_{mm}(\beta) &=& -j(j+1)d^j_{mm}(\beta) + m^2\sec^2(\beta/2)d^j_{mm}(\beta)\nonumber\\
&-& \cot\beta\frac{d}{d\beta}d^j_{mm}(\beta).
\label{3.1a13}
\end{eqnarray}
 Putting $\beta =2i\xi$, we obtain
\begin{equation}
\frac{d^2\Delta}{d\xi^2} = 4j(j+1)\Delta - 4\frac{{m^2}\Delta}{{\cosh^2}\xi}
- 2\coth2\xi\frac{d\Delta}{d\xi}.
\label{2.3.19a1} 
\end{equation}
Using Eq. $(\ref{2.3.17})$ this can be expressed as
\begin{equation}
\frac{d^2\Delta}{d\xi^2} = 4j(j+1)\Delta - 
4\frac{{m^2}\Delta}{{\cosh^2}\xi} - \Gamma\frac{\cosh2\xi}
{{\cosh^2}\xi}. 
\label{2.3.20}
\end{equation}
Thus, we have obtained all the necessary averages
of the pseudo-spin operators over $|\Psi_{m}\rangle$
to calculate $\Delta{J_{x^\prime}}^2$ and 
$\Delta{J_{y^\prime}}^2$.
                                                                                
Now,                                                                               from Eqs. $(\ref{2.3.3})$, $(\ref{2.3.4})$ and 
$(\ref{2.3.12})$ it is evident 
that the mean pseudo-spin vector $\langle\hat{{\mathbf J}}\rangle$ is not
along the $z$-axis and lies
in the $z-x$ plane making an angle say $\theta_{1}$ with the $z$-axis. Therefore, we 
align the vector $\langle\hat{{\mathbf J}}\rangle$ along the $z^\prime$-axis by
performing a rotation as below.
\begin{eqnarray}
\hat{J}_{x^\prime} &=& \hat{J}_{x}\cos\theta_{1} - 
\hat{J}_{z}\sin\theta_{1}
\label{2.4.1}\\ 
\hat{J}_{y^\prime} &=& \hat{J}_{y}
 \label{2.4.2}\\
\hat{J}_{z^\prime} &=& \hat{J}_{x}\sin\theta_{1} + 
\hat{J}_{z}\cos\theta_{1} 
\label{2.4.3}
\end{eqnarray}

with
\begin{equation}
\tan\theta_{1} = \frac{\langle{\hat{J}_{x}}\rangle}
{\langle{\hat{J}_{z}}\rangle}.
\label{2.4.4}
\end{equation}
This rotation makes 
$\langle\hat{J}_{x^\prime}\rangle = 0$ and
since
$\langle{\hat{J}_{y}}\rangle$ is already zero as given in 
Eq. $(\ref{2.3.4})$, the
vector $\langle\hat{\mathbf J}\rangle$ is now along the $z^\prime$-axis. We now observe
the variances in $J_{x^\prime}$ and 
$J_{y^\prime}$.
As $\hat{J}_{y^\prime} = \hat{J}_{y}$, hence the
variance in $J_{y^\prime}$ is the
same as ${J}_{y}$ i.e.
\begin{eqnarray}
\Delta J_{y^\prime} &=& 
\sqrt{\langle\hat{J}_{y^\prime}^2\rangle 
- {\langle{\hat{J}_{y^\prime}\rangle}^2}}\nonumber\\
&=& \sqrt{\langle{{\hat{J}_{y}}}^2\rangle - 
{\langle{\hat{J}_{y}}\rangle^2}}\nonumber\\
&=& \Delta{J_{y}}.
\label{2.4.4a}
\end{eqnarray}                                                                      
As $\langle \hat{J}_y \rangle = 0$ by Eq. $(\ref{2.3.4})$,
therefore,
\begin{equation}
\Delta{J_{y}}^2 = \langle{\hat{J}_{y}}^2\rangle = \frac{1}{4\Delta}\coth\xi
\frac{d\Delta}{d\xi} = \Delta{J_{y^\prime}}^2
\label{2.4.5}
\end{equation} 
where we have used Eq. $(\ref{2.3.16})$. Using Eq. 
 (\ref{2.3.17}), we get
\begin{equation}
\Delta J_{y^\prime}^2 = \frac{\Gamma}{4\Delta}.
\label{3.3}
\end{equation}

On the other hand since 
$\langle \hat{J}_{x^\prime} \rangle = 0$,
 the square of the variance in 
$J_{x^\prime}$ is given as
\begin{equation}
\Delta J_{x^\prime}^2 = 
\langle \hat{J}_{x^\prime}^2\rangle.
\label{2.4.5newa1}
\end{equation} 
Therefore, using Eq. $(\ref{2.4.1})$,  we get                                                                     
\begin{eqnarray}
\Delta J_{x^\prime}^2 &=& \langle\hat{J}_{x}^2\rangle \cos^{2}\theta_{1} +
\langle{\hat{J}_{z}}^{2}\rangle \sin^{2}\theta_{1}\nonumber\\
&-& \langle{\hat{J}_{x}}\hat{J}_{z} +
\hat{J}_{z}\hat{J}_{x}\rangle \sin\theta_{1}\cos\theta_{1}.
\label{2.4.6}
\end{eqnarray}                                                                                
Using Eqs. $(\ref{2.3.13})$, $(\ref{2.3.14})$, 
$(\ref{2.3.15})$, $(\ref{2.3.17})$, (\ref{2.3.20}) and $(\ref{2.4.4})$
in Eq. (\ref{2.4.6}), we obtain
\begin{eqnarray}
\Delta J_{x^\prime}^2 &=& {\Bigg{[}\frac{m^2}{\cosh^2\xi} +
 \frac{\tanh^2\xi}{4}{\Big{(}\frac{\Gamma}{\Delta}\Big{)}^2} \Bigg{]}}^{-1}\Bigg{[}{\Big{(}\frac{\tanh^2\xi}{4}\Big{)}}^2\nonumber\\&\times&{\Big{(}\frac{\Gamma}{\Delta}\Big{)}}^3 + \frac{j(j+1)m^2}{\cosh^2\xi} - \frac{m^2}{4\cosh^4\xi}
\frac{\Gamma}{\Delta}\Bigg{]}\nonumber\\
&-& \frac{m^2}{\cosh^2\xi}.
\label{3.2}
\end{eqnarray}
                                                                            
It is to be noted that since $\frac{\Gamma}{\Delta}$ is symmetric under $m\longrightarrow -m$,
 which can be verified from the Eqs. $(\ref{3.4})$, 
$(\ref{2.3.18})$ and $(\ref{2.3.19})$, the quantity
$\Delta J_{x^\prime}^2$ and $\Delta J_{y^\prime}^2$ are also symmetric under $m\longrightarrow -m$. For $m=\pm{j}$
we have
\begin{equation}
\Delta J_{x^\prime}^2 = \Delta J_{y^\prime}^2 = \frac{j}{2}.
\label{2.4.8}
\end{equation}
The value of $|\langle\hat{\bf J}\rangle|$ for the state
$|\Psi_m\rangle$ is obtained using Eqs. (\ref{2.3.3}), (\ref{2.3.4}), (\ref{2.3.12}) and (\ref{2.3.17}) as
\begin{eqnarray}
|\langle\hat{\bf J}\rangle| &=& \sqrt{\langle\hat{J}_x\rangle^2 + \langle\hat{J}_y\rangle^2 + \langle\hat{J}_
z\rangle^2}\nonumber\\ 
&=& {\Big{[}\frac{m^2}{\cosh^2\xi}  
+ \frac{\tanh^2\xi}{4}{\Big{(}\frac{\Gamma}{\Delta}\Big{)}^2} \Big{]}}^{1/2}.
\label{3.6}
\end{eqnarray}

Using Eqs. (\ref{3.3}) and (\ref{3.2}) in Eq. (\ref{2.12}), we can calculate the value of the quantum 
entanglement parameter $E$, numerically, for various values of the radiation field parameter $\xi$, and for various numbers $N$ of two-level atoms in the system. The numerical value of $E$ can be taken as the amount of quantum entanglement present among the atoms. If $E$ is zero, then there is no quantum 
entanglement among the atoms and the atoms are completely
uncorrelated. 

\begin{figure}
\begin{center}
\includegraphics[width=6cm, angle=-90]{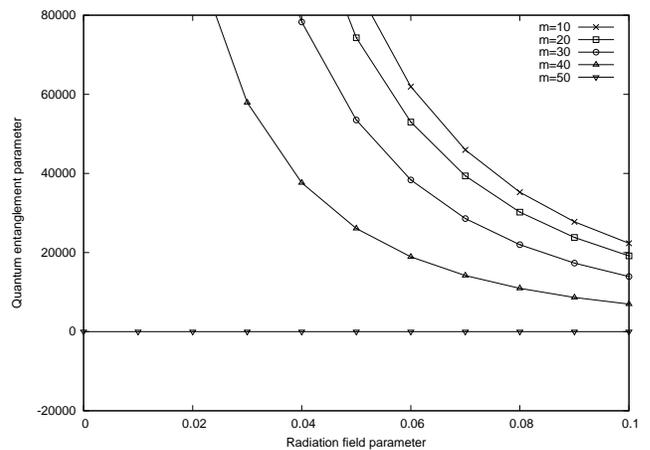}
\caption{Variation of the quantum entanglement parameter $E$ as a function of the radiation field parameter $\xi$ for $N=100$. $E$ and $\xi$ are plotted on
the vertical and horizontal axes respectively.}
\end{center}
\label {fig1}
\end{figure}

Using Eq. (\ref{3.6}) in Eq. (\ref{2.14}), and experimentally measuring the spectroscopic squeezing parameters 
$\xi_{R_{x}}$ and $\xi_{R_{y}}$, we can calculate the quantum entanglement parameter $E$ from Eq. (\ref{2.14}).

We, now, show in Fig. 1, the variation of the quantum entanglement parameter $E$ for one hundred two-level atoms in interaction with the squeezed vacuum state of the radiation field. We plot graphs for various values of $m$. The non-zero
value of $E$ indicates the presence of quantum entanglement and the numerical value of $E$ can be taken as the amount of
quantum entanglement present among the atoms of the system.

We observe from Fig. 1 that the quantum entanglement parameter $E$ starts from a very high value when the radiation field parameter $\xi = 0$, and sharply decreases, as $\xi$ increases. We can find that $E$ is very close to $0$,
when $\xi$ is close to $3$. We also
observe that, for a fixed value of $\xi$, the amount of quantum entanglement increases with decrease in $m$ values. We note that, $E$ is always zero for $m= 50$, which is the maximum value of $m$ for $N = 100$.
In Table I, we present the values of $E$ with $\xi$ for different values of $m$, when $N = 100$.
\begin{table}[ht]
\caption{Values of $E$ with $\xi$ for different values of 
$m$, with $N = 100$}
\centering
\begin{tabular}{c c c c c c c c c}
\hline\hline
m & $\xi$ & $E$ & $\xi$ & $E$ & $\xi$ & $E$ & $\xi$ & $E$\\ [0.5ex]
10 & 0 & 1440000 & 0.1 & 22357.14 & 0.2 & 4914.34 & 3 & 0.0142
\\
20 & 0 & 1102500 & 0.1 & 19162.98 & 0.2 & 4189.47 & 3 & 0.0109\\
30 & 0 & 640000 & 0.1 & 13943.33 & 0.2 & 3013.98 & 3 & 0.0063\\
40 & 0 & 202500 & 0.1 & 6965.37 & 0.2 & 1470.5 & 3 & 0.002\\
50 & 0 & 0 & 0.1 & 0 & 0.2 & 0 & 3 & 0\\ [1 ex]
\hline
\end{tabular}
\label{table:quant}
\end{table}

\begin{figure}
\begin{center}
\includegraphics[width=6cm, angle = -90]{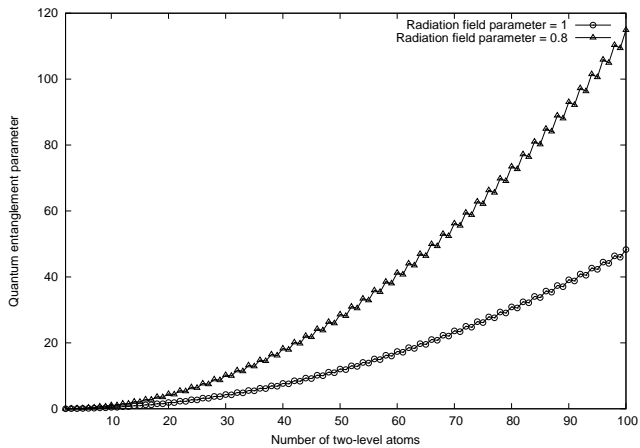}
\caption{Variation of the quantum entanglement parameter 
$E$ with the number of atoms  $N$ for two fixed values of 
$\xi$, which are $\xi = 0.8$ and $\xi = 1$. Here we have taken $m = 1$. $E$ and $N$ are plotted on
the vertical and horizontal axes respectively.}
\end{center}
\label {fig2}
\end{figure}
The numerical values of $E$, given in Table I, can be taken as the measure of quantum entanglement for the corresponding values of $\xi$, for hundred atoms.

\begin{figure}
\begin{center}
\includegraphics[width=6cm, angle = -90]{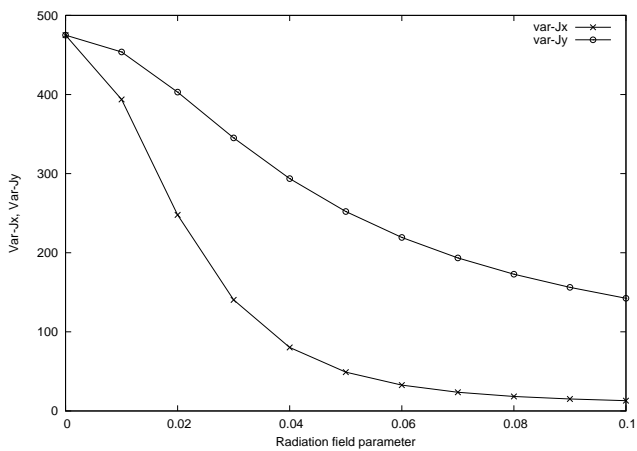}
\caption{Variation of ${\Delta J_{x^\prime}}^2$ and ${\Delta J_{y^\prime}}^2$ with radiation field parameter $\xi$ for
 $m = 40$. ${\Delta J_{x^\prime}}^2$ and 
${\Delta J_{y^\prime}}^2$ have been shown as $Var-Jx$ and $Var-Jy$ respectively in the figure.}
\end{center}
\label {fig3}
\end{figure}

In Fig. 2 we graphically show the variation of the quantum entanglement parameter $E$ with the number of atoms varying
from $N = 2$ to $N = 100$ for fixed values of $\xi$. Here, we have taken $m = 1$. The two
curves correspond to two fixed values of $\xi$, which are
$\xi =0.8$ and $\xi = 1$. We observe that $E$ increases with
$N$, and also the values of $E$ are higher at $\xi = 0.8$ than those at $\xi = 1$. We can also see from Eqs. (\ref{3.3}), (\ref{3.2}), and $(\ref{3.6})$, and from Eqs. (\ref{2.12}) to (\ref{2.14}), that the amount of quantum entanglement present in the system is independent of the sign of $m$, that is, $E$ is invariant under the transformation
$m \longrightarrow -m$.

In Fig. 3 we show the variation of ${\Delta J_{x^\prime}}^2$ and ${\Delta J_{y^\prime}}^2$ with the radiation field parameter. In the figure, these quantities have been shown as
$Var-Jx$ and $Var-Jy$ respectively. We observe that both the quantities ${\Delta J_{x^\prime}}^2$ and 
${\Delta J_{y^\prime}}^2$ have values greater than 100 in the
range from $\xi = 0$ to $\xi = 0.03$. Thus, in this range of 
$\xi$, the quantum state $|\Psi_m\rangle$ shows no spin squeezing at all, whereas, we observe from Fig. 1 that in this range of $\xi$, the quantum entanglement parameter
$E$ has a very high value, implying the presence of quantum entanglement among the two level-atoms. Thus, presence of quantum entanglement among the atoms does not ensure spin squeezing, as defined in Ref. \cite{Kitagawa}, of the system. 

\section{IV. SUMMARY AND CONCLUSION}

We quantify the amount of quantum entanglement present in a system of $N$ two-level atoms in interaction with the squeezed vacuum state of the electromagnetic field.
The method, that we use to quantify the amount of quantum entanglement present in such system, is to calculate the numerical value of the mean squared deviations of the quantum fluctuations in $x^\prime$ and $y^\prime$ quadratures of the considered quantum state from the corresponding quantum fluctuations of an atomic coherent state or coherent spin state (CSS). We use this method, because,
we have expressed the composite quantum fluctuations in $x^\prime$ and $y^\prime$ quadratures of $N$ two-level atoms as an algebraic sum of the corresponding quantum fluctuations of the $N$ individual constituent atoms and the quantum correlation terms among them. We have found that these multiparticle quantum correlation terms $CORRX$ and $CORRY$ are equal to the deviations of the quantum fluctuations in $x^\prime$ and $y^\prime$ quadratures respectively, of the corresponding composite quantum state from those of an atomic coherent state or coherent spin state (CSS), which is a completely separable state \cite{Ram}, \cite{Kitagawa}. We have also seen that the multiparticle quantum correlation terms $CORRX$ and $CORRY$ are made up of all possible bipartite (two-atoms) quantum correlations among the $N$ atoms. Now, since these multiparticle correlation terms may be positive or negative, we take the mean of the squares of $CORRX$ and $CORRY$ as a measure of the amount of multiparticle quantum entanglement present in the system of $N$ two-level atoms. This mean squared value of $CORRX$ and $CORRY$ is equal to the mean squared value of the deviations of the quantum fluctuations in $x^\prime$ and $y^\prime$ quadratures of the considered quantum state from the corresponding quantum fluctuations of an atomic coherent state. We, thus, quantify the amount of quantum entanglement present in the system of $N$ two-level atoms.   
 
 We graphically show the variation of the multiparticle quantum entanglement with the radiation field parameter for one hundred such two-level atoms, for different values of $m$. We note that the amount of quantum entanglement for a fixed value of the radiation field parameter, increases with the decrease in the value of $m$. We also note that, the amount of quantum entanglement present in the system remains invariant under the transformation $m \longrightarrow -m$. We show graphically, the continuous variation of the amount of quantum entanglement present in the system, as we increase continuously the number of atoms in the system from $N = 2$ to 
$N = 100$, for two fixed values of the radiation field parameter. We note that, the amount of quantum entanglement for any number of atoms $N$ is higher for lesser value of the radiation field parameter. We also note that, the amount of quantum entanglement in the system of atoms increases, as we increase the number of atoms in the assembly.

We also show that in certain range of the radiation field parameter, the system shows
no spin squeezing, as defined in Ref. \cite{Kitagawa}, but shows high quantum entanglement. 

Our method can also be applied to the system studied by Felicetti et al., \cite{Felicetti}, where they propose a superconducting circuit architecture for multipartite entanglement generation. It can also be applied to the system studied in Ref. \cite{Clark}, in which they propose a scheme to controllably entangle the internal states of two atoms trapped in a high-finesse optical cavity by employing quantum reservoir engineering. Our method can also be applied to the system used in Ref. \cite{Matthias}, where they studied the dynamics of the entanglement of formation in a quantum-measurement model, consisted of a four-level atom harmonically bound in a three-dimensional trap.

We hope that our method of quantification of the amount of quantum entanglement in multiparticle systems may produce some new insight into the subject. 

\section{ACKNOWLEDGEMENT}
I am grateful to late Binayak Dutta Roy and Nilakantha Nayak for useful discussions with them.

\end{document}